\documentclass[prl,showpacs,preprintnumbers,amsmath,amssymb,endfloats,
letterpaper]{revtex4}

\usepackage{graphicx}
\usepackage{dcolumn}
\usepackage{bm}
\usepackage{epstopdf}
\usepackage{latexsym}
\newcommand{\be}{\begin{equation}}
\newcommand{\ee}{\end{equation}}
\newcommand{\bea}{\begin{eqnarray}}
\newcommand{\eea}{\end{eqnarray}}

\newcommand{\req}[1]{Eq.~(\ref{#1})}

\begin{document}

\title{Orbital Response of Evanescent Cooper Pairs\\  in Paramagnetically Limited Al Films}

\author{X. S. Wu and P. W. Adams}

\affiliation{Department of Physics and Astronomy\\ Louisiana State University\\ Baton Rouge, Louisiana, 70803}%
\author{G. Catelani}
\affiliation{Physics Department\\Columbia University\\New York, NY 10027}

\date{\today}

\begin{abstract}
We report a detailed study of the pairing resonance via tunneling density of states in ultra-thin superconducting Al films in supercritical magnetic fields. Particular emphasis is placed on effects of the perpendicular component of the magnetic field on the resonance energy and magnitude. Though the resonance is broadened and attenuated by $H_\bot$ as expected, its energy is shifted upward linearly with $H_\bot$.  Extension of the original theory of the resonance to include strong perpendicular fields shows that at sufficiently large $H_\bot$ the overlap of the broadened resonance tail with the underlying degenerate Fermi sea alters the spectral distribution of the resonance via the exclusion principle.  This leads to the shift of the the resonance feature to higher energy.   
\end{abstract}

\maketitle

Though superconductivity is one of the most ubiquitous and extensively studied phenomenon in condensed matter physics, little is known about how an attractive electron-electron interaction mode, whether mediated by phonons or some more exotic coupling, evolves from a non-coherent fluctuation channel to a full blown superconducting instability.   Because the formation of stable Cooper pairs usually coincides with a precipitous collapse into a macroscopic quantum condensate, it is difficult to extract properties of individual Cooper pairs in the superconducting phase.  Likewise there are few systems in which a Cooper pair precursor can even be identified in the normal state \cite{PseudoGap,Precursor1,Precursor2}, much less studied. Here we show that spin-paramagnetically limited superconducting Al films provide an unique opportunity to directly probe the energetics and dynamics of the Cooper pair precursor near the threshold of a well-defined superconducting instability associated with the first-order spin-paramagnetic (SP) transition.  Using electron tunneling spectroscopy, we demonstrate that the precursor, which is manifest as a finite bias anomaly in the tunneling spectrum, exhibits an unexpected energy shift associated with the orbital response to the perpendicular component of the applied magnetic field.

Although the spin-paramagnetic transition was conjectured over forty years ago \cite{Clogston1962,Chandrasekhar1962}, the dynamics and complete phase diagram of the transition have only been mapped out in the last decade \cite{Wu1994,Butko1999a,Adams1998,Rogachev2005,Ralph1997,Braun1997}. This is, in part, due to the fact that the first-order nature of the transition and the corresponding tricritical point ($T_{Tri}\sim600$ mK in Al) are extremely sensitive to and quickly extinguished by spin-orbit (SO) scattering. Consequently, many of the salient features of the SP transition such as hysteresis \cite{Wu1994}, state memory \cite{Butko1999a}, avalanches \cite{Wu1995a}, and the finite bias tunneling anomaly \cite{Wu1996} require a true spin-singlet ground state. This precludes the observation of such effects in high atomic number superconductors in which there is a strong spin-orbit scattering rate, such as there is in Pb and Sn, for instance \cite{Tedrow1982,Adams2004}. Indeed, there are only two well documented spin-singlet superconductors, Be and Al.  

In practice, the SP transition can be probed by applying a magnetic field parallel to the surface of a film that is sufficiently thin ($< 5$ nm) so as to suppress the Meissner response.  The field thus fully penetrates the film, and the superconducting phase remains unperturbed until the Zeeman splitting becomes of the order of the superconducting gap.  At that point a first order phase transition to the paramagnetic normal state occurs \cite{Clogston1962,Chandrasekhar1962}.  The spin symmetry of ultra-light BCS superconductors is not only manifested in the SP transition itself but in the paramagnetic normal state as well. In particular, an unexpected pairing resonance (PR), clearly associated with virtual Cooper pair formation, was recently discovered in the paramagnetic normal state density of states (DOS) spectrum of Al and Be films \cite{Wu1996,Adams2000}. The resonance can be observed in high field tunneling by virtue of the fact that it is a spin-singlet mode riding on top of a paramagnetic background. In general terms its energy is the sum of the Zeeman energy of the anti-aligned electron in the virtual Cooper pair minus the effective binding energy gained by forming the pair.  A rigorous nonperturbative analysis by Altshuler \textit{et.\ al} 
\cite{Aleiner1997,Kee1998} showed that in parallel field the resonance occurs at an energy that is universal for 0D(grain), 1D(wire), and 2D(film) systems,
\be\label{Parallel}
E_+=(E_z+\Omega)/2,
\ee
where $\Omega=\sqrt{E_z^2-\Delta_o^2}$, $E_z=g_L\mu_BH$ is the Zeeman energy, $H$ the magnetic field in Tesla, $g_L$ the Land\'e $g$-factor, $\mu_B$ the Bohr magneton, and $\Delta_o$ the zero temperature, zero field, superconducting energy gap.   Though  \req{Parallel} was found to be in reasonably good agreement with parallel field experiments \cite{Butko1999}, a comprehensive comparison between experiment and theory was never made.  In particular, though it is known that the resonance becomes more stable with decreasing film conductance, its evolution in the limit that the dimensionless conductance $g=h/e^2R\rightarrow 1$, where $R$ is the sheet resistance of the film, remains an open question. Furthermore, the theory was developed under the assumption that the Zeeman energy dominates the spectrum.  This assumption will be violated in a tilted field if the Cooperon cyclotron energy $\hbar\Omega_H=4DeH_\bot\sim E_z$, where $D$ is the electron diffusivity and $H_\bot$ is the perpendicular component of the field.   In this Letter, we take advantage of improved sample quality to investigate the PR in low $g$ films in the presence of large perpendicular magnetic fields. We show that contrary to expectation, the PR energy in tilted field is not solely a function of the Zeeman splitting, but, in fact, has a contribution that is proportional to the perpendicular component of the field.

The films were grown by e-beam deposition of 99.999\% Al onto fire polished glass microscope slides held at 84 K. The depositions were made at a rate of $\sim0.1$ nm/s in a typical vacuum $P<3\times10^{-7}$ Torr. A series of films with thicknesses ranging from 2 to 2.9 nm had a dimensionless normal state conductance that ranged from $g=5.6$ to $213$ at 100 mK. After deposition, the films were exposed to the atmosphere for 0.5-4 hours in order to allow a thin native oxide layer to form. Then a 9-nm thick Al counterelectrode was deposited onto the film with the oxide serving as the tunneling barrier. The counterelectrode had a parallel critical field of $\sim$2.7 T due to its relatively large thickness, which is to be compared with $H_{c\parallel}\sim6$ T for the films. The junction area was about 1 mm$\times$1 mm, while the junction resistance ranged from 10-100 k$\Omega$ depending on exposure time and other factors.  Only junctions with resistances much greater than that of the films were used.  Measurements of resistance and tunneling were carried out on an Oxford dilution refrigerator using a standard ac four-probe technique. Magnetic fields of up to 9 T were applied using a superconducting solenoid. A mechanical rotator was employed to orient the sample \textit{in situ} with a precision of $\sim0.1^{\circ}$.

\begin{figure}
\includegraphics[width=6in]{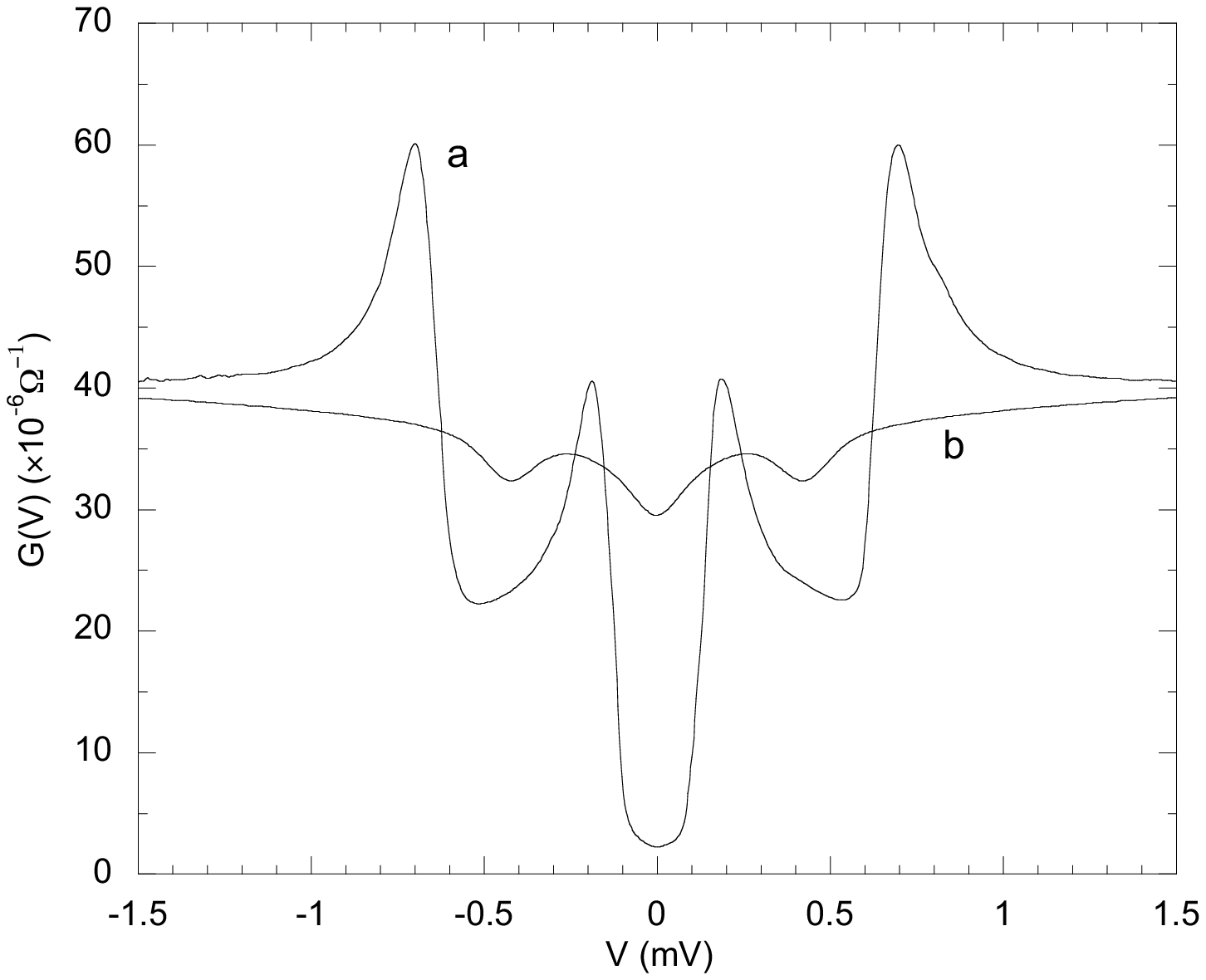}
\caption{\label{TypicalCR} Tunneling conductance of a $g=12$ film at 60 mK in superconducting and normal states, where parallel fields are 4.5 T (a) and 5.7 T (b), respectively. In the superconducting state, the Zeeman split of BCS DOS manifests itself as two peaks of tunneling conductance on either side of $V=0$. }
\end{figure}

Two typical 60-mK tunneling spectra are shown in Fig.\ \ref{TypicalCR} in which the parallel field is above and below $H_{c\parallel}$. All of the tunneling data presented in this paper were made with the counter-electrode in the normal state.  In this configuration the low temperature tunnel conductance is proportional to the DOS of the film \cite{Tinkham1980}. In the superconducting state, the BCS DOS is Zeeman-split by the applied magnetic field as can be seen in the 4.5-T curve of Fig.\ \ref{TypicalCR}. As one further increases the field, a first-order transition to the normal state occurs near $H_{c\parallel}=\Delta_o/\sqrt{g_L}\mu_B$ \cite{Clogston1962,Chandrasekhar1962,Wu1995b}. The normal state spectrum, represented by the 5.7 T curve in Fig.\ \ref{TypicalCR}, displays a logarithmic depletion of states near the Fermi energy, commonly known as the zero bias anomaly (ZBA). It is now well established that the logarithmic ZBA is associated with electron-electron interactions in two-dimensional disordered systems \cite{Altshuler1985}. The finite bias anomalies in the 5.7-T curve are due to the PR and represent a window into the processes that ultimately lead to the formation of stable Cooper pairs at lower fields.

\begin{figure}
\includegraphics[width=6in]{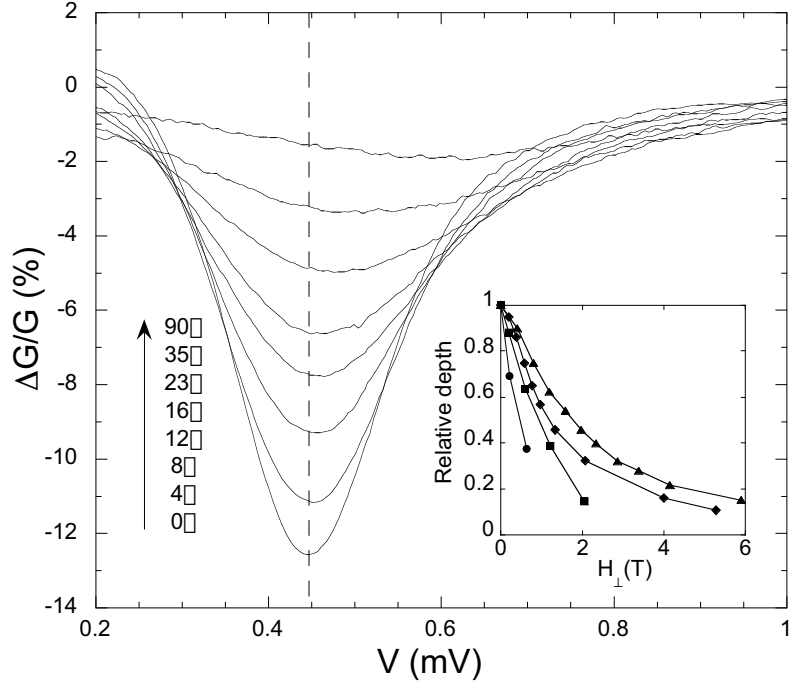}
\caption{\label{TiltedField} Pairing resonance at several field orientations for a $g=5.6$ film at 60 mK in a field of 5.9 T.  $\theta=0$ corresponds to field parallel to the film plane.  The vertical dashed line marks the position of the parallel field resonance.  Inset: the relative depth of the PR as a function of the perpendicular component of magnetic field for several film conductances: Circle $g=58$, Square $g=26$, Diamond $g=12$, Triangle $g=5.6$.}
\end{figure}

In Fig.\ \ref{TiltedField} we plot the resonance feature in a $g=5.6$ sample at a variety of tilt angles $\theta$, where $\theta$ is the angle between the field and the plane of the film. The field was held constant in this data set so that the perpendicular field component was $H_{\bot}=H\sin\theta$. Two features of the data should be noted. First, though the PR is attenuated and broadened by $H_{\bot}$, it survives to much larger tilt angles in the lowest conductance films. This is clear in the inset of Fig.\ \ref{TiltedField} where the relative depth of the PR is plotted as a function of $H_{\bot}$ for films of different $g$. Second, the resonance moves to higher energy with increasing tilt angle. 

\begin{figure}
\includegraphics[width=6in]{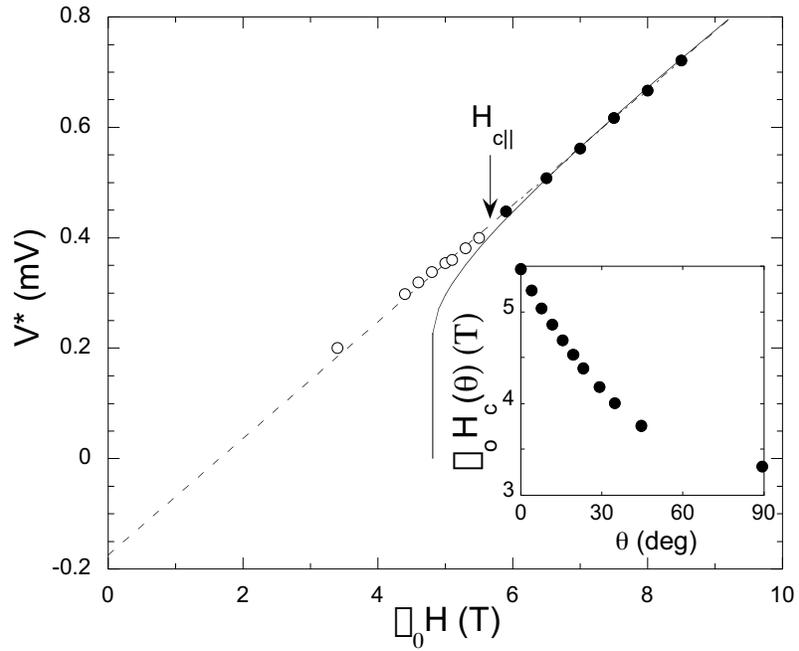}
\caption{\label{Vstar-H} The position of the PR as a function of magnetic field for the $g=5.6$ film at 60 mK. The vertical arrow marks the parallel critical field. Solid symbols correspond to parallel field.  The open symbols were taken at $\theta=8^{\circ},12^{\circ},16^{\circ},19^{\circ},23^{\circ},29^{\circ},45^{\circ}, {\rm and}\ 90^{\circ}$ (from right to left). The dashed line is provided as a guide to the eye. The solid line is a fit of \req{Parallel} to the parallel field data. The inset shows the critical field as a function of tilt angle.}
\end{figure}

The measurement of PR in tilted magnetic fields allows one not only to test the effect of perpendicular field, but also extend the measurements to fields well below the spin-paramagnetic critical field, $H_{c\parallel}$. Upon rotation of a few degrees out of parallel the first-order spin-paramagnetic transition becomes second-order and the critical field rapidly begins to decrease until finally reaching $H_{c2}$ at $\theta=90^\circ$. This latter behavior can be seen in the inset of Fig.\ \ref{Vstar-H}. In the main panel of Fig.~\ref{Vstar-H}, we plot the resonance voltage as a function of applied field for the $g=5.6$ sample. The solid symbols represent parallel field data which, of course, must terminate at $H_{c\parallel}$ due to the fact that the film undergoes a first-order transition to the superconducting phase at this field. The solid line represents a fit to the parallel field data using \req{Parallel}, where only the g-factor was varied, with a best fit value $g_L=1.61$ \cite{Butko1999}. The gap value used in \req{Parallel} was taken from measurements in the superconducting phase, $\Delta_o/e=0.43$ mV. The open symbols in Fig.~\ref{Vstar-H} represent measurements made just above the critical field at finite tilt angles up to $\theta=90^\circ$. Note that the tilted field data are linear in H well below $H_{c\parallel}$ and that this behavior is inconsistent with Eq.\ \ref{Parallel}. 

\begin{figure}
\includegraphics[width=6in]{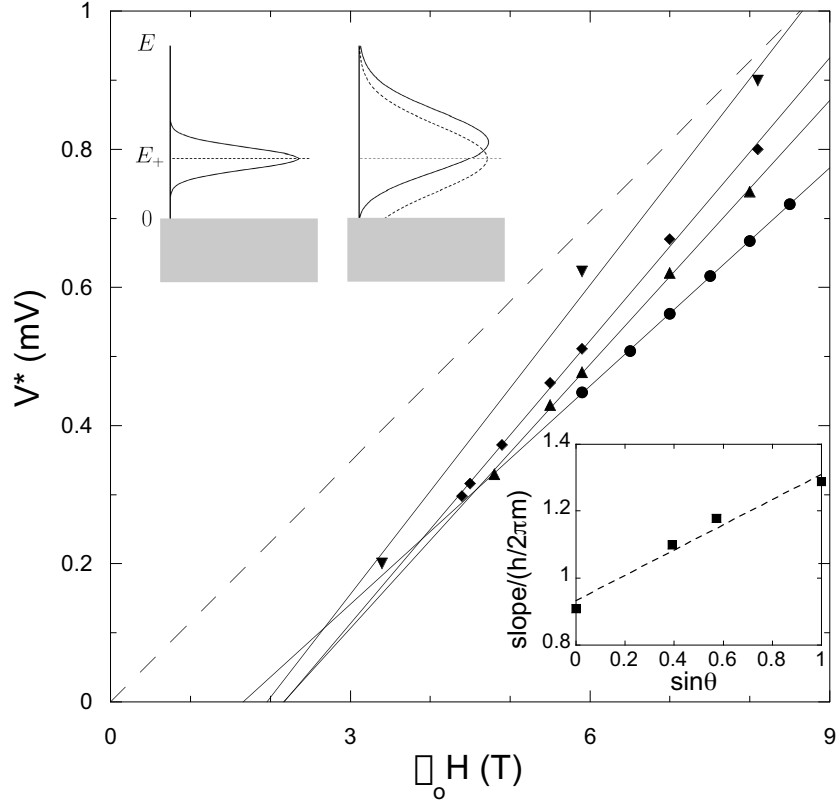}
\caption{\label{Slope-Angle} The position of the PR as a function of magnetic field for the $g=5.6$ film at 60 mK at several different field orientations: circle $\theta=0^{\circ}$, up triangle $\theta=23^{\circ}$, diamond $\theta=35^{\circ}$, and down triangle $\theta=90^\circ$. The solid lines are linear fits to the data. For comparison, the Zeeman voltage $V_z=E_z/e$ is plotted as a dashed line, taking $g_L=2$.  Upper inset: Width and position of the resonance for small (left) and large (right) perpendicular field as explained in the text.  Lower inset: the fitted slope $dV^*/dH$ in units of $\hbar/m$ from the data in the main panel plotted as a function of $\sin\theta$. The dashed line is a linear fit to the data and has slope $0.38$.}
\end{figure}

The open symbol data points in Fig.~\ref{Vstar-H} were obtained by changing both the field magnitude and the tilt angle. A more straightforward characterization of the field dependence of the resonance can be obtained by varying the field at a preset tilt angle. Shown in Fig.\ \ref{Slope-Angle} is the field dependence of the resonance energy at tilt angles $\theta=0^\circ$, $23^\circ$, $35^\circ$, and $90^\circ$.  Not only is the linearity of the parallel field dependence preserved in tilted field, but the slope of the field dependence grows with increasing tilt angle.   

We have extended the original analysis of the resonance to include the case where the perpendicular magnetic field is such that the Cooperon cyclotron energy is much larger than the characteristic width of the resonance
\be\label{limit}
\hbar\Omega_H=4eDH \sin \theta \gg W_2=\frac{\Delta_o^2}{4g\Omega}.
\ee
The above limit violates an approximation used in the derivation of \req{Parallel} that requires the Cooperon momentum transfer to be small.  In the parallel field case, the transition from the superconducting phase to the paramagnetic phase occurs at the Clogston limit $E_z\sim\sqrt{g_L}\Delta_o$; consequently the evanescent pair energy $\Omega$ lies well above the Fermi energy, $\Omega\gtrsim \Delta_o$.  Assuming that the momentum transfer is small, i.e. $W_2 \ll \Delta_o$, one can neglect the presence of the degenerate Fermi sea below the pair, since $E_+ > \Omega$.  In the presence of a perpendicular 
magnetic field, however, Landau quantization of the electron motion leads to a momentum transfer (in energy units)
of the order of $\hbar\Omega_H$, and when this becomes sufficiently large the exclusion principle can no longer be ignored. When taking into account the 
Fermi sea, the position of the resonance feature is shifted to higher energies, as
we explain by the diagram in the upper inset of Fig.~\ref{Slope-Angle}. On the left
we show qualitatively the anomaly in the density of states for $\hbar\Omega_H\ll W_2$: 
the anomaly is centered at $E_+$ and its width is given by $W_2$;  since $W_2\ll E_+$, the presence of the Fermi sea (gray area) is not felt 
by the excitations determining the anomaly.
On the right, the dashed line shows the anomaly as it would be without the 
Fermi sea for $\hbar\Omega_H\gg W_2$; a finite fraction of excitations 
contributing to the anomaly  would now be located below the Fermi
energy. The solid line represents the ``true'' shape of the anomaly. As the 
exclusion principle suppresses the contribution of the excitations below
the Fermi energy, the profile of the anomaly becomes asymmetric and the
observed minimum is shifted to higher energies; moreover 
we expect that the larger $\Omega_H$ is, the larger the shift becomes, in 
agreement with the quantitative result given below.

As will be described in detail elsewhere \cite{Catelani2005}, the total correction to the density of states can be written as the sum of two terms,
$\delta \nu = \delta \nu_{1}+ \delta \nu_2$, where the first is the perpendicular field result previously obtained in Ref.~\onlinecite{Kee1998} and the second is the new contribution of the present analysis. We reemphasize that the minimum of $\delta\nu_{1}$ is always centered at $E_+$, for any tilting angle. While the full expression for $\delta\nu_2$ is somewhat lengthy and beyond the scope of this Letter -- it is obtained by
a perturbative (one-loop) calculation, justified by the condition in \req{limit}, which includes subleading contributions -- it is this term that shifts the apparent position of the resonance to higher energy. By linearizing the energy derivative of $\delta \nu$, we obtain the following approximate formula for the position $V^*$ of the resonance in tilted field:
\be
V^* =  E_+/e +\alpha4DH\sin\theta\, , \quad 
\alpha = -\frac{1}{\pi^5} \psi''\left(\frac{1}{2}\right) \simeq 0.055
\label{Vstar}
\ee
where $\psi$ is the digamma function.  This generalizes the result of Ref.~\onlinecite{Kee1998} and reduces to it in the limit $H_\bot=0$.  The $E_+$ term in \req{Vstar} is well approximated by a linear field dependence, see the dashed line Fig.~\ref{Vstar-H}, and has no angle dependence.  Consequently, the contribution of the orbital term can be extracted from the linear fits to the tilted field data in Fig.~\ref{Slope-Angle} (solid lines).  The slopes values obtained from these fits are plotted as a function of $\sin\theta$ in the lower inset of Fig.~\ref{Slope-Angle}.  The dashed line in the inset has a slope of $0.38\hbar/m$ which by \req{Vstar} should be equal to $\alpha4D$.  This gives a corresponding diffusivity $D\sim1.7\hbar/m$ which is in excellent agreement with the value of the diffusivity obtained from the film conductance $D=g/4\pi\hbar\nu t\sim1.4\hbar/m$, where $\nu$ is the renormalized density of states and $t\sim1$ nm is the film thickness after accounting for surface oxidation.

In conclusion, we have studied the effect of perpendicular field on pairing resonance in the low temperature paramagnetic normal state of ultra-thin Al films.  Tilting the magnetic field away from parallel orientation broadens the resonance thereby producing an overlap with the underlying Fermi sea.  The exclusion principle effectively produces a ``boundary condition'' that truncates  the low side of the resonance tail, resulting in an apparent upward shift in the resonance energy.   Since the resonance represents a microscopic window into the fundamental quantum processes by which a stable Cooper pair is formed out of a specific interaction channel, a more comprehensive understanding of its dynamics has far reaching implications.  In particular, further studies in granular Al films with significantly lower conductance may reveal a  dimensionality crossover from 2D to 0D as $g\rightarrow1$, as well as illuminate the role of {\it e-e} interaction effects in the evolution of the resonance.  

We gratefully acknowledge enlightening discussions with Igor Aleiner, Ilya Vekhter, Dana Browne, and Nandini Trivedi. This work was supported by the National Science Foundation under Grant DMR 02-04871.

\bibliographystyle{apsrev}
\bibliography{PairingResonance}

\newpage

\newpage

\newpage

\newpage

\end{document}